\documentclass[aps,prl,showpacs,preprintnumbers,twocolumn,nofootinbib,longbibliography]{revtex4-1}
\usepackage{geometry}                
\expandafter\let\csname equation*\endcsname\relax
\expandafter\let\csname endequation*\endcsname\relax

\usepackage[normalem]{ulem}
\usepackage{graphicx}
\usepackage{amssymb}
\usepackage{dsfont}
\usepackage{epstopdf}
\usepackage{color}
\usepackage{mathtools}
\usepackage{cancel}
\usepackage{hyperref}
\usepackage{chemformula}

\usepackage{physics}

\definecolor{red}{rgb}{1,0,0}
\definecolor{blue}{rgb}{0,0,1}
\definecolor{green}{rgb}{0,0.75,0.5}
\definecolor{black}{rgb}{0,0,0}

\newcommand{\p}{\partial}
\newcommand{\eq}[1]{\begin{align}#1\end{align}}
\newcommand{\eqs}[1]{\begin{align*}#1\end{align*}}

\newcommand{\half}{\mbox{$\frac{1}{2}$}}
\newcommand{\hhalf}{\frac{1}{2}}

\usepackage{calrsfs}
\renewcommand{\pv}{\vec{p}}

\newcommand{\xv}{\vec{x}}
\newcommand{\xiv}{\vec{\xi \;}}

\newcommand{\nv}{{\vec n}}
\newcommand{\piv}{\vec{\pi}}

\newcommand{\PP}{\mathbb{P}}

\begin{document}
\title{Noise equals control}
\author{Eric De~\!Giuli}
\date{\today}                     
\affiliation{Department of Physics, Toronto Metropolitan University, Canada}

\begin{abstract}
Stochastic systems have a control-theoretic interpretation in which noise plays the role of control. In the weak-noise limit, relevant at low temperatures or in large populations, this leads to a precise mathematical mapping:  the most probable trajectory between two states minimizes an action functional and corresponds to an optimal control strategy. In Langevin dynamics, the noise term itself serves as the control. For general Markov jump processes, such as chemical reaction networks or electronic circuits, we use the Doi-Peliti formalism to identify the `response' (or `momentum') field $\pi$ as the control variable. This resolves a long-standing interpretational problem in the field-theoretic description of stochastic systems: although $\pi$ evolves backward in time, it has a clear physical role as the control that steers the system along rare trajectories. This implies that Nature is constantly sampling control strategies. We illustrate the mapping on multistable chemical reaction networks, systems with unstable fixed points, and specifically on stochastic resonance and Brownian ratchets. The noise-control mapping justifies agential descriptions of these phenomena,
and builds intuition for otherwise puzzling phenomena of stochastic systems:  why probabilities are generically non-smooth functions of state out of thermal equilibrium; why biological mechanisms can work better in the presence of noise; and how agential behavior emerges naturally without recourse to mysticism. 
\end{abstract}
\maketitle


\section{Introduction}

Systems biology faces the monumental task of synthesizing vast amounts of molecular facts into a cohesive whole \cite{Kitano02,Kitano04,Bialek12}. 
While it is recognized that stochasticity is ubiquitous in the cell, with noise playing a role in gene expression, differentiation, and switching (see reviews \cite{Kaern05,Raj08,Bialek12,Sanchez13,Tsimring14,Levchenko14,Bressloff17,Ilan23}), it is still commonly argued that the cell functions {\it in spite} of the noisy cellular environment. Biologically relevant lower bounds on copy number fluctuations add to the puzzle \cite{Lestas10,Hilfinger16,Yan19}. Even the overall program of systems biology at the cellular level has been summarized as an attempt to answer 
three key questions \cite{Dehghani24}: (i) where are the control switches? (ii) how to manage the need to reconfigure? (iii) how to harness noise rather than succumb to it? No holistic perspective is available that unites these three questions. 
  Here we show that (i) and (iii) are in fact two sides of the same coin: noise plays the mathematical role of control, and is to be exploited, not overcome. This resolves dissonance in the literature and may help to build a unified understanding of the cell as a system under persistent endogenous control.

{ 
We will show in particular that a noise-control mapping provides the solution to two conceptual problems:

First, for stochastic systems evolving in continuous time, it has been known since the 1970's that the complete statistics of all observables can be obtained from a path-integral approach, known as Martin-Siggia-Rose-De-Dominicis-Janssen \cite{msr,De-Dominicis76,Janssen76} if obtained from a Langevin equation, and Doi-Peliti \cite{Doi76,Peliti85} \footnote{More correctly it should be called Doi-Zel'dovich-Grassberger-Goldenfeld-Peliti \cite{Doi76,ZelDovich78,Grassberger78,Goldenfeld84,Peliti85} } if obtained from a Markov jump process, like a chemical reaction network (CRN). In either case, for a system whose original stochastic description has $N$ fields $\{ n_j(t) \}$, the path integral involves 2N fields, a doubling of degrees of freedom. As recalled later, it can be ensured by a change of variable if necessary that $N$ of the fields correspond to the original physical fields $\{ n_j(t) \}$. 
The remaining $N$ fields, which we will write as $\{ \pi_j(t) \}$, relate to the noise, and have been called `momentum', `response', `tilt', or `quantum' fields, but what is their precise physical interpretation? 
 A difficulty is that their evolution equations 
 must be solved {\it backward} in time. This seems to make it impossible to assign a physical meaning to them, and thus to the trajectories, if causality is to be preserved.

Second, at the scale of the organism, biology is suffused with a language of agency, defined simply as goal-directed behavior \cite{Ball23} \footnote{For a more expansive view in relation to biology, see \cite{Newman25}.}: the predator tries to catch prey; the bird builds a nest for its future young; and so on. Since the advent of  video microscopy, this language has crept to smaller scales, for example in the famous video of David Rogers showing a neutrophil `chasing' a bacterium through a field of red blood cells (Movie 3.1, \cite{Phillips12}). Recent experiments on limb regeneration show that cells can `decide' to grow new limbs from environmental cues, suggesting a form of agency at the cellular level \cite{Pezzulo16}. How agency arises from matter is then a fundamental problem. Where is the dividing line between a reductionist description and an agential one, if one indeed exists? If everything in biology has an ultimate mechanistic description, how does this not conflict with agential descriptions? Unless {\it all} use of teleological language is dropped or usurped from biology, this is a conceptual problem of dualism, called the `agency-body' problem in \cite{Levin25}. In modern discourse the tacit solution to the agency-body problem is that agency is emergent at some level in the description of biological processes. While the failure of a purely reductionist approach to capture emergent behavior is well-recognized in general \cite{Laughlin00}, and for biology in particular \cite{Nicholson19}, and while emergence has a definition in renormalization group theory \cite{Chaikin00,Tauber12}, if this is the general explanation for agency in biology it requires a specific mechanism, which has not been found. 

We show below that both problems are simultaneously solved by a widened perspective that is  mathematically explicit and general.

\subsection{Roadmap}

We first outline the general connection between noise and control, and explain how the present approach differs from previous work. We then discuss this in a Hamiltonian formalism, and solve the first conceptual problem outlined above. This is illustrated on a double-well potential. We then introduce to a physics audience several deep results from control theory. We extend the discussion from Langevin dynamics to full Markov jump processes, further cementing the solution to the first conceptual problem. After discussing the noise-control mapping in general multistable systems, and connecting to stochastic thermodynamics, we discuss the appearance of control-stabilized fixed points. We then discuss the phenomena of stochastic resonance and Brownian ratchets from the point of view of the noise-control mapping. We illustrate some of the paradoxical features of the emergent dynamics by way of analogy, discuss prospective global optimization principles, and finally conclude. 

MATLAB codes to reproduce numerical results are available at \cite{data_noise}.
}

\section{Noise equals control}

Consider a state space of $N$ abundances $n_j$, called species, well-mixed in a volume $\Omega$. We focus on the transition probability $\PP(\nv_f | \nv_0)$ to go from $\nv(0) = \nv_0$ to $\nv(t_f) = \nv_f$ in time $t_f$. We begin with a Langevin equation
\eq{ \label{eq1}
\p_t \nv + \vec F(\nv) = \hat B(\nv) \cdot \xiv
}
where the noise $\xiv$ has $P$ elements, with $\langle \xi_a \rangle=0$ and $\langle \xi_a(t)\xi_b(t') \rangle = \delta_{ab}\delta(t-t')$. { $\vec F$ is the vector-valued drift field, and the matrix $\hat B$ multiplies the noise. The drift field can be decomposed into potential and non-potential parts \cite{Fang19}. Here and in the following the dynamics may contain explicit time dependence, but we suppress it in the notation. Write $\sim$ for `asymptotic to' in the large $\Omega$ limit, i.e. $A \sim B$ if $\lim_{\Omega \to \infty} A/B$ is finite. If $\nv$ represents the mesoscale population dynamics of a microscopic system, such as a chemical reaction network, or an ecosystem, then typically $\nv$ and $F$ are extensive, $\nv \sim F \sim \Omega$, while $B$ is sub-extensive, for example $B \sim \sqrt {\Omega}$ if the noise arises from population noise, i.e. the underlying stochasticity of reaction events. (Later we will also address other sources of noise). To see this scaling explicitly it is convenient to pass to a representation in terms of intensive variables (if $\nv$ is a particle count then $\nv/\Omega$ is a concentration) (e.g. \cite{Falasco25}). As a result of these scalings, 
 the macroscopic limit $\Omega \to \infty$ is a small-noise limit in which the noise term carries a factor $1/\sqrt{\Omega}$ relative to the other terms. Our analysis also applies to small-noise limits obtained for other reasons, like low temperature. In any small-noise limit} Eq.\ref{eq1} is dominated by solutions minimizing the action 
\eq{ \label{eq2}
S = \int dt \; L(\nv,\xiv) 
}
with $L(\nv,\xiv) = \half \xiv^2$, and subject to boundary conditions $\nv(t_f) = \nv_f$ and $\nv(0) = \nv_0$.
The minimal solution (henceforth called instanton) determines both the most likely trajectory and the fluctuations around it.  

{ Separately, consider a system with state $\xv(t)$ and dynamics $\p_t \xv = \vec f(\xv(t)) + \hat g(\xv(t))\cdot \vec u(t)$, where $\vec f$ is a vector and $\hat g$ is a matrix of the appropriate size,  involving an affine control $\vec u(t)$.} A control problem is to choose $\vec u(t)$ to fulfill an objective, such as making the system go between two states in a given time \cite{Sontag13,Glad18,Bechhoefer21}. An {\it optimal control }problem is to do so while minimizing a cost function $C[\xv,\vec u]$. As an example, consider a rocket with states defined by height and vertical velocity, placed in a vector $\xv(t)$, and engine thrust $u(t)$ as control; the Goddard problem is to maximize the height obtained for a given mass of fuel \cite{Glad18}.  

{ 
We emphasize that a control problem is structurally different from the type of problem normally considered in physics: the control is left free, rather than subject to dynamics imposed by Nature. This freedom gives the formalism flexibility, for the control can then be tailored to the experimental situation. For example, if the control needs to be specified explicitly in advance, then it cannot depend on the state $\vec x(t)$, and is called ``open-loop", whereas if the control is a function of the state, then it is a ``feedback controller", or ``closed-loop" control. 

Control theory makes precise a very general notion of agency: the agent is whoever or whatever selects the control. The goal is the control objective, such as getting the system from $\vec x_0$ to $\vec x_f$. If the control achieves the goal and is not null, 
 then by definition the system exhibits goal-directed behavior. 

This specification goes beyond optimization theory in that it defines the decision made or action taken by the agent. For example, free energy is minimized in thermal equilibrium, but classical thermodynamics is completely silent on how this minimization is achieved. There is no specified agent and no explicit control.  
}

Remarkably, the optimization program to minimize Eq.\eqref{eq2} subject to Eq.\eqref{eq1} is mathematically identical to an optimal control problem, where $\xiv$ plays the role of affine control variable. 

{
The connection between action-minimizing trajectories of a stochastic system and optimal controls has been noticed in the large deviations literature \cite{Fleming77,Fleming06,Chetrite15,Chetrite15b,Grafke19,Jack20}, and considered as a mathematical mapping useful for numerical computation of rare events \cite{Heymann08,Grafke19}. Recent work \cite{Chetrite15,Chetrite15b} considers a mapping at the level of stochastic processes: on one hand, transition probabilities of the original stochastic system, and on the other hand a `controlled' process in which rare trajectories of the original process become typical ones of the controlled process. This mapping does not exist when the underlying landscape is non-convex \cite{Chetrite15,Chetrite15b}, which will be the main regime of interest in this work. 

We propose instead the following: in all stochastic systems, the noise can be interpreted as a control variable; no separate control system needs to be studied. Either in an experiment in which we observe a single trajectory, or in a theoretical path integral formulation in which we average over trajectories to evaluate observables, the control is no longer free, as it would be in control theory. So it is not an external control in that sense; we propose to call it endogenous control when the noise is a byproduct of the natural dynamics (as in population/demographic noise); we will also briefly consider the case when noise is environmental, where one might call it exogenous control. In what follows the source of the noise is not crucial: what is important is that Nature automatically samples different controls.

The objective that the control attempts to achieve is to get the system from initial to final state; in the macroscopic limit, it attempts to do so while minimizing the action. This clarifies that the macroscopic limit, in which control is optimal, is not necessary for the noise-control mapping. The latter can be universally applied. This distinguishes the present approach from previous work. 



}


The interpretation of noise as control is the main point of this work. After generalizing beyond Langevin dynamics to general Markov jump processes, like chemical reaction networks (CRNs), we illustrate the noise-control mapping in various multistable systems with and without time-dependent forcing. 
 In all cases the noise-control mapping is a Rosetta stone between a language of mechanics and a language of agency \cite{Ball23}: every statement about the role of noise in a macroscopic stochastic system can be translated to a statement about optimal control. 




 For example, the {\it lac} operon is a well-studied bistable system \cite{Eldar10}. Its transition from the `off' state to the `on' state depends on a rare event (coincident dissociation of an inhibitor at two locations) \cite{Choi08}. We suggest that the noise causing this transition can and should be considered as a control. 
 
{ Conversely, control problems can be solved by noisy processes \cite{Kappen05,Kappen05b}. In Appendix 1 we illustrate this for the Goddard problem. Although this is not necessarily an efficient method to numerically find instantons, Nature automatically samples noisy processes, and therefore can solve control problems. }

\begin{figure*}[th!]
\centering
\includegraphics[width=\textwidth]{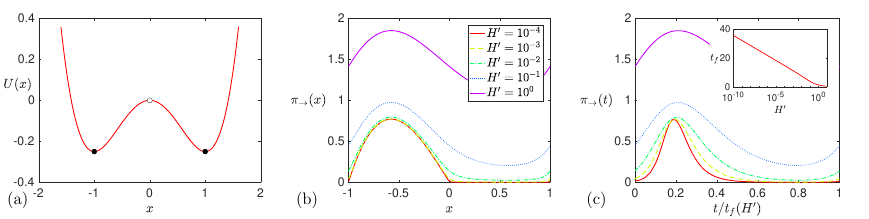}
\caption{ (a) Double-well potential $U(x) = -\half a x^2 + \frac{1}{4} b x^4$, with $a=b=1$; the stable minima $x = \pm 1$ and unstable saddle $x=0$ are marked. (b) Control to go from $x=-1$ to $x=+1$ as a function of $x$ at indicated $H'=H/\Omega$. (c) Same controls, illustrated as a function of time, normalized by the total time for the transition (see inset). }\label{well1}
\end{figure*}

Before further unpacking this mapping, we stress that it is distinct from the external control of stochastic systems \cite{Touchette00,Cao09,Sagawa12,Blaber23}. It is a statement about the objective that the stochastic system is {\it already achieving}. We return to this point in the Discussion.

\section{Hamiltonian formulation} 

For practical computations it is useful to pass to a Hamiltonian formulation. In a path integral formulation, the field equation 
Eq.\eqref{eq1} is introduced with a Lagrange multiplier $\piv(t)$ so that 
\eq{ 
S & \to  \int dt \; \left[ \half \xiv^2 + \piv \cdot (\p_t \nv + \vec F(\nv) - B \cdot \xiv) \right]  \notag \\
& \to \int dt \; \left[ \piv \cdot \p_t \nv - H(\nv,\piv) \right] \label{eq3}
}
where we integrated out $\xiv$, implying $\xiv = B^T \cdot \piv$. { The relationship between $\piv$ and $\xiv$ can be understood by the example of a reaction network, which as recalled later has the same action Eq.\eqref{eq3}: in a CRN, particle numbers only change when a reaction occurs, so the noise is fundamentally associated with reactions, and will be $M$-dimensional if there are $M$ reactions, like $\xiv$ here. But we can also represent the noise, for an appropriate $B$ matrix, by its effect on the particle counts, and this is what is represented by $\piv$. In general, }up to normalization, $\piv$ is the noise expanded on the species, and therefore a control expanded on the species. The Hamiltonian is $H(\nv,\piv) = -\piv \cdot \vec F(\nv) + \half \piv \cdot B \cdot B^T \cdot \piv$. 

In the macroscopic limit {both $n$ and $H$ are extensive, $n \sim H \sim \Omega$, which again is most easily seen by a change of variable to concentrations: the limit is defined by finite values of $\lim_{\Omega \to \infty} \vec F(\Omega \vec x)/\Omega$ and $\lim_{\Omega \to \infty} B(\Omega \vec x)/\Omega^{1/2}$ at finite $\vec x = \vec n/\Omega$. Then in the macroscopic limit $\Omega \to \infty$ the path integral is dominated by a saddle-point trajectory and the Gaussian fluctuations around it. The saddle-point equations take a Hamiltonian form 
\begin{subequations} \label{ham} \eq{ 
\p_t \nv & = + \nabla_\pi H \\
\p_t \piv & = -\nabla_n H 
} \end{subequations} 
where e.g. $(\nabla_\pi H)_j = \p H/\p \pi_j$. Note that $\piv=0$ is always a solution, and reduces Eq.\eqref{ham} to the deterministic equation at zero noise. Note that $\piv$ here is not a stochastic variable; it is deterministic, subject to Eq.\eqref{ham}.

Eq.\eqref{ham} has a formal similarity to classical mechanics, and some properties of the latter follow immediately, such as conservation of $H$ along trajectories whenever $H$ does not carry explicit time-dependence: $dH/dt = \nabla_\nv H \cdot \p_t \nv + \nabla_\pi H \cdot \p_t \piv = 0$. However despite this similarity, Eq.\eqref{ham} is fundamentally different, because it is solved with boundary conditions $\nv(t_f) = \nv_f$ and $\nv(0) = \nv_0$, rather than two initial conditions as in classical mechanics. 
In particular, the $\piv$ equation is solved running time backwards. }
The noise-control mapping explains why: in order to reach a final state $\nv(t_f)=\nv_f$, we must choose the control $\piv(t)$ appropriately. Whenever $\nv_f$ is not the state to which deterministic dynamics would lead, this information must be propagated backwards from the final time. 

{
As a simple example consider a double-well potential $U(x) = -\half a x^2 + \frac{1}{4} b x^4$, pictured in Fig.\ref{well1}a, and $F(n) = \Omega \p_x U(x)$. The Hamiltonian is, taking $B=\Omega^{1/2}$,
\eq{
H(x,\pi)/\Omega = \pi (ax-bx^3) + \half \pi^2
}
and the instanton equations are
\begin{subequations} \label{dblwelleq} \eq{
\p_t x & = a x - bx^3 + \pi \\
\p_t \pi & = - \pi (a - 3b x^2)
} \end{subequations}
Setting $\pi=0$ and $\p_t x=0$ to find the deterministic fixed points, we get the two stable minima, $x_\pm = \pm \sqrt{a/b}$ and the unstable saddle $x=0$. These all correspond to $H=0$. Now let us look for time-varying instantons. It is convenient to use conservation of $H$, yielding a family of solutions $\pi(x) = bx^3-ax \pm \sqrt{(a-bx^2)^2 x^2 + 2 H'}$ parameterized by $H'=H/\Omega$. Consider instantons going from $x_-$ to $x_+$. The total time taken is
\eqs{
t_f = \int dt = \int \frac{dx}{\p_t x} = \int \frac{dx}{\sqrt{(a-bx^2)^2 x^2 + 2 H'}}
}
From here it is clear that larger positive $H$ corresponds to faster transitions. As $H \to 0^+$ the integrand develops divergences both at the saddle and at the stable fixed points: in the limiting case $H=0^+$ the system starts arbitrarily close to the fixed point and escapes in infinite time. The ideal case $H=0$ is instructive: we have either $\pi=0$ or $\pi=2x(bx^2-a)$ and in fact $\pi(x) = 2x(bx^2-a)$ on the `uphill' part of the trajectory ($x<0$) and $\pi(x)=0$ on the `downhill' part. The instanton is not smooth at the saddle $x=0$.

( There are also trajectories with $H<0$ but these do not correspond to hops; they will be considered later. )

The control is illustrated both as a function of $x$, in Fig.\ref{well1}b, and as a function of $t$, in Fig.\ref{well1}c, at indicated $H'$, for $a=b=1$. The asymmetry of $\pi(x)$, and the fact that $\pi>0$, are clear indications that it is agential and acausal. The controls to go in the reverse direction can be obtained by symmetry. 

If this example is reinterpreted as the dynamics of an overdamped particle, then the agential interpretation is even more transparent. If $x$ corresponds to the position of an overdamped particle, then $\pi_\rightarrow$ corresponds intuitively to the magnitude of force needed to bring the particle from $x_-$ to $x_+$ in a given time, while minimizing $\int dt \;\pi^2$. For example at $H=0^+$ the particle is pushed to the saddle and then relaxes freely to $x_+$. The interpretation of $\pi$ as agential control is natural. 

In the weak-noise limit, Nature automatically selects these trajectories for hops. Although this may appear trivial in this low-dimensional example, since any path from $x_-$ to $x_+$ must necessarily go through all intermediate $x$, the same principle holds in high-dimensional systems with many attractors, where the nature of optimal trajectories is much less obvious.

We note that double-well systems also appear in out-of-equilibrium chemical reaction networks, such as the Schl{\"o}gl model \cite{Schlogl72,Vellela09}, or the genetic toggle switch \cite{Roma05,Heymann08}. In these cases there is no spatial symmetry but the controls have a qualitatively similar form.

}


In general, in the macroscopic limit the leading behavior of a transition probability is $\PP(\nv_f | \nv_0) \propto e^{-S_*}$ where $S_*$ is $S$ evaluated on the instanton $(\nv_*(t),\piv_*(t))$ from $\nv_0$ to $\nv_f$ over the interval $t \in (0,t_f)$. The corrections, including the prefactor, are also fixed by the instanton. As discussed in Appendix 2, they depend on a matrix $Q(t)$ characterizing the curvature around the instanton. Its inverse has a control-theoretic interpretation as a feedback matrix, relating deviations from $\nv_*$ to the control that brings the system there.

\section{Immediate results from control theory} 

The most obvious use of the noise-control mapping is to take concepts, intuitions, and theorems from optimal control theory \cite{Sontag13} and see what they say about stochastic systems in the weak-noise limit. 


A famous result is the Pontryagin maximum principle (PMP), which states that, for Eqs.(\ref{eq1},\ref{eq2}) and general $L$, the optimal control must have a Hamiltonian form, equivalent to Eq.\eqref{ham} (see Appendix 3), but obtained without path integral manipulations. Importantly, the PMP does not assume that the optimal control is smooth, and in fact it often has a pointwise behavior with discontinuities, particularly when the control is limited to a finite or closed set. Such functions are approximated arbitrarily well by Brownian $\xiv(t)$ integrated over in a standard Langevin approach. This is shown explicitly for the Goddard problem in Appendix 1, whose optimal control is discontinuous. We will see later that non-smooth controls are generic in nonequilibrium physical systems. 

{ We now turn to a series of deep results concerning internal models \cite{Ashby58,Conant70,Francis76,Isidori02,Sontag03}. Consider a control system in which we add an external disturbance $\vec f_{ext}(t)$:
\eq{
\p_t \xv = \vec f(\xv(t)) + \hat g(\xv(t))\cdot \vec u(t) + \vec f_{ext}(t),
}
where the controls are restricted to some set $U$. A natural control problem is to find a control $u(t) \in U$ such that the state can be maintained in a desirable family of states $X$ despite the disturbance, e.g. $\vec x(t_f) \in X$. If the disturbance belongs to a family $F = \{ f_{ext} \}$, then the system rejects $F$ if the control problem can be solved for any $f_{ext} \in F$. Ashby's law of requisite variety \cite{Ashby58} states that to do so, it is necessary that the control be of large enough dimension: $\text{dim}(U) \geq \text{dim}(F) - \text{dim}(X)$ \footnote{It can also be stated in information-theoretic terms \cite{Conant70}.}. Generalizations of this result are known collectively as the Internal model principle \cite{Conant70,Francis76,Isidori02,Sontag03}, which further state that for a system to reject $F$, the system 
must contain a controller that models all disturbances in $F$, and feeds back into the system to counterract the disturbance. For example, if $f_{ext}$ is generated by a dynamical system $dz/dt = s(z)$ (called `exosystem') with $f_{ext}=q(z)$, then the controller must internally reproduce solutions to $dz/dt=s(z)$ within its dynamics. This is a model of the disturbance. For example, constant $f_{ext}$ corresponds to $s(z)=0$, and the controller must contain an integrator, which solves $dz/dt=0$. }


This principle has been used to motivate biologically relevant control architectures, particularly for `robust perfect adaptation', which is rejection of constant disturbances \cite{Briat16,Xiao18,Briat23}. 
It is accomplished by integral feedback using explicit regulator species. It was found empirically that noise stabilizes the proposed mechanisms, as expected by the present approach. {Indeed, adding noise and interpreting it as control makes disturbance rejection easier, since it increases $\text{dim}(U)$ in Ashby's law, and makes it easier to build an internal model, for the same reason. } 

\section{Extension to Markov jump processes} 

The identification of noise with control is not limited to Langevin dynamics. For Markov jump processes, like CRNs, the full counting statistics are derived from a Doi-Peliti field theory \cite{Doi76,Peliti85} 
built from Doi's Hamiltonian formulation of the jump process \cite{Doi76}. { As mentioned above, such a path integral has a doubling of degrees of freedom. Usually it is introduced with two fields $\phi_j$ and $\phi^*_j$ whose component-wise product $\phi_j\phi^*_j$ corresponds to the original physical field $n_j$. By a Cole-Hopf transformation \cite{Andreanov06,Kamenev02,Smith11,De-Giuli22b} we make $\nv$ one of the fields, and $\pi_j = \log \phi^*_j$ the other. }  The action then has the form Eq.\eqref{eq3}, where the Hamiltonian $H(\nv,\piv)$ is now a general function of $\nv$ and the momenta $\piv$ satisfying $H(\nv,0)=0$, which enforces conservation of probability (for pedagogical reviews see \cite{Kamenev02,Weber17,Lazarescu19,De-Giuli22b,Falasco25}). 


In the macroscopic limit Eq.\eqref{ham} continues to apply. Due to $H(\nv,0)=0$, $\piv =0$ is always a solution, and gives back the deterministic trajectories. However these will only be compatible with the boundary conditions if $\nv_f$ is the state reached from $\nv_0$ under deterministic dynamics. To reach other states, active control is necessary, i.e. $\piv \neq 0$. 
 If the Hamiltonian is expanded in small $\piv$, then the Langevin equation is recovered at quadratic order, with $\xiv = B^T \cdot \piv$, so the previous analysis applies and $\piv$ is still a control.  

Away from this small $\piv$ regime, the dynamics goes beyond Langevin, but it is still determined by Eq.\eqref{ham} in the macroscopic limit. One may wonder if $\piv$ can still be identified as a control in this regime. This is so, most easily seen as follows: consider $L(\nv,\piv) = \piv \cdot \nabla_\pi H - H$ and the control problem to minimize $\int L$ with $\piv$ as control. Applying the PMP one finds after a few steps (see Appendix 3) the same Hamilton equations, and the objective becomes exactly $S=\int L$. Therefore $\piv$ remains a control. 

Finally, away from the macroscopic limit one has to deal with full path integrals over $\nv(t),\piv(t)$. Each path contributes proportional to $e^{-S}$. Although instantons will have the largest contribution to the transition probability, other paths will contribute as well; $\piv$ is still a control, but control is not necessarily optimal. 

Thus the `response' or `momentum' field $\piv$, whose interpretation has always been obscure, can universally be interpreted as a control. This is important because it gives the path ensemble a concrete physical interpretation: the existence of the ensemble means that, in principle, we need only average over paths to evaluate observables; the interpretation of $\piv$ as control means that different paths correspond to different control strategies. Nature is always attempting control.

\section{Multistable systems at long times} { Dissipative systems show a dichotomy between two types of dynamics: relaxation towards attractors, and escapes from attractor to attractor. Since }biological systems need both robustness and variety, they are expected to be multistable at many scales, 
 with attractors playing the role of functional states \cite{Waddington14,Smith16,Qian16,Smith20}. In CRNs, a multistable landscape can only exist out of equilibrium. 

{ In this section we discuss general features of the noise-control mapping in the multistable case, in particular when there are many attractors, or when the underlying space is high-dimensional. We focus on the long-time limit where sharp statements can be made. 

To see why the long-time limit introduces new features, consider again the double-well example of Fig.\ref{well1}. If we ask for the probability $\PP(+1|-1, t)$ to start at $x=-1$ and end at $x=+1$ in a short time $t$, then this will be dominated by single-hop trajectories as discussed above, and $\PP(+1|-1, t) \approx t \kappa_{-1,+1}$ where $\kappa_{-1,+1}$ is the transition rate for a hop from $-1$ to $+1$. However over longer time-scales, the system will eventually hop back to $-1$, even if it ends at $+1$. In the infinite time limit $t \to \infty$, the system can hop arbitrarily many times from well to well, and it forgets memory of the initial condition. Stationarity of the long-time jump process between attractors implies $\PP(+1|-1, t \to \infty) = 1/(1 + \kappa_{1,-1}/\kappa_{-1,1})$, which depends on both `forward' and `backward' transition rates. In systems with many attractors, these stationary probabilities do not admit a simple expression (see \cite{De-Giuli23} ). We now discuss this case in general, following \cite{Falasco25}. In a first reading, this section can be omitted.
}

The objective in the multistable case is still maximization of $\PP(\nv_f|\nv_0)$, but in general it is a two-step process. If the system has multiple basins of attraction $\{\gamma\}$, then locally one can only determine the relaxations to the fixed-point $\gamma$ along with action-minimizing uphill trajectories to saddles. To construct globally optimal trajectories one has to patch together these trajectories at saddles. 

\begin{figure}[t!]
\centering
\includegraphics[width=0.95\columnwidth]{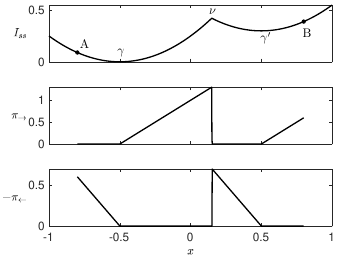}
\caption{ Illustration of quasipotential $I_{ss}$ in a bistable system (a), along with controls $\pi(x)$ to go from A to B (b) and from B to A (c). }\label{fig_Iss}
\end{figure}

For time-independent rates define the quasipotential $I_{ss}(\nv) = \lim_{\Omega \to \infty} (-\log \PP(\nv))/\Omega$ achieved in the long time limit $t \to \infty$, which loses memory of initial conditions. 
Away from singularities it is fixed (up to a constant) by $H(\nv,\nabla I_{ss})=0$ in the macroscopic limit, and $I_{ss}$ is a Liapunov function for the deterministic dynamics. In fact within each basin of attraction $\gamma$ one constructs the local quasipotential $I^{(\gamma)}_{ss}$. At each $\nv$ the global quasipotential is fixed by
\eq{ \label{Iss}
I_{ss}(\nv) = \min_\gamma (I_{ss}^{(\gamma)}(\nv)+\alpha_\gamma) - \min_\gamma \alpha_\gamma ,
}
where the constants $\{ \alpha_\gamma \}$ are fixed by a jump process over attractors. The sum $-(\alpha_\gamma+I_{ss}^{(\gamma)}(\nv))$ is interpreted as the log-probability to be in attractor $\gamma$, plus the log-probability to reach $\nv$ from the fixed point in $\gamma$, all divided by $\Omega$. Thus the optimization problem involves both a local component, fixing $I_{ss}^{(\gamma)}$, and a global component, fixing the $\{ \alpha_\gamma \}$. See Fig.\ref{fig_Iss} (top) for an illustration, { where the plotted coordinate $x$ could be one coordinate of many in a high-dimensional space.}

It follows from Eq.\eqref{Iss} that $I_{ss}$ is not smooth at the boundaries between basins. We call these boundaries `saddles' although they may not be located at the saddles expected from a naive analysis. It was shown by Graham and T\'el in a series of works that this non-smooth behavior is generic whenever the system is not in detailed balance, except in the special case when the $H=0$ manifold is integrable \cite{Graham84,Graham84a,Graham85,Graham86} ( see review \cite{Baek15}). 

This phenomenon, which may initially appear exotic, has a natural interpretation from the control point of view. First we note that on uphill instantons we have $\piv~=~\nabla I_{ss}$~\footnote{This follows from conservation of $H$ in the time-independent case, along with the fact that instantons leave fixed points (where $H=0$).}, while on downhill instantons we have $\piv=0$. This allows us to define a control $\piv_\eta(\nv)$ for the instantonic path $\eta$ from one state to another, where we must take the appropriate branch depending on whether we go uphill $(\nv \cdot \nabla I_{ss} > 0)$ or downhill $(\nv \cdot \nabla I_{ss} < 0)$. This is illustrated in Fig.\ref{fig_Iss}bc for a schematic bistable system, for paths going from A to B and from B to A (In this illustrative example, the controls were found from the quasipotential). The interpretation of $\piv$ is clear: the system needs to be steered into the desired attractor. After passing through the saddle, the control can be turned off, as the system will relax freely to the fixed point. (To go further uphill beyond the fixed point, the control needs to be turned back on.) So { $\nabla I_{ss}$ reflects different dynamical processes on either side of a saddle, and it }has no need to be continuous at saddles, so that $I_{ss}$ can have kinks, as observed. Moreover it is natural to describe these paths as agential. 

This path-dependent $\piv_\eta(\nv)$ has a quantitative role since the transition rate $\kappa_{\gamma\gamma'}^\nu$ from attractors $\gamma$ to $\gamma'$ via the saddle $\nu$ can be written \cite{Falasco25}, in the macroscopic limit, as
\eq{ \label{kappa}
\log \kappa_{\gamma\gamma'}^\nu & = -\Omega [ I_{ss}^{(\gamma)}(\nv_\nu) - I_{ss}^{(\gamma)}(\nv_\gamma) ] \notag \\
& = -\Omega \int_\gamma^\nu d\nv \cdot \piv 
} 
where the line integral is along the instanton up to the saddle. These rates in turn fix the constants $\{ \alpha_\gamma \}$ appearing in Eq.\eqref{Iss} \footnote{In terms of the stationary probabilities $\mathcal{P}_\mu$ of the jump process over attractors, $\alpha_\mu = -\log (\mathcal{P}_\mu)/\Omega$.}.  The quantity $1/\kappa_{\gamma\gamma'}^\nu$ is the mean-first-passage time, so that 
the timescale associated to a path $\eta$ is $\tau_\eta = e^{\Omega \int_\eta d\nv \cdot \piv_\eta}$. Control is paid for in time: {transitions that require more control will only occur (typically) over longer time scales.}

 
 


\begin{figure}[t!]
\centering
\includegraphics[width=.65\columnwidth]{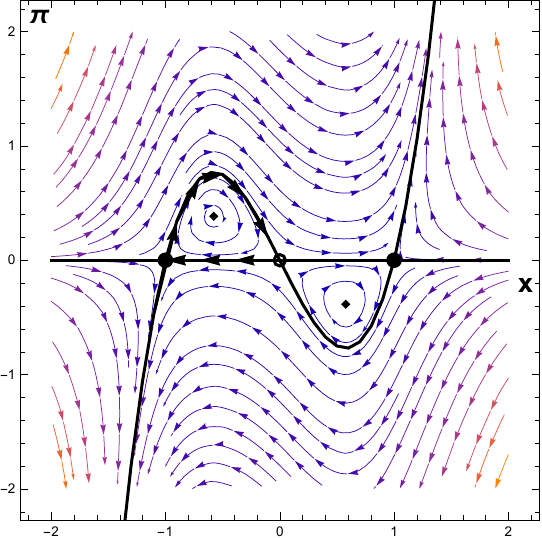}
\caption{ Phase portrait of double-well system in $x-\pi$ space. The $H=0$ manifold is shown in bold. This includes both deterministic trajectories at $\pi=0$ and the uphill instantons. The (uphill) instanton from $x=-1$ to $x=0$, and the relaxation trajectory from $x=0$ back to $x=-1$ are indicated with arrows. The cycles around the control-stabilized fixed points are clear.}\label{phaseportrait}
\end{figure}

\subsection{Stochastic Thermodynamics}

So far we have not imposed any conditions arising from thermodynamics, but they are readily incorporated with the stochastic thermodynamics (ST) formalism \cite{Seifert12,Freitas22,Falasco25}. Under the standard assumptions of ST, and in the macroscopic limit, the transition rates between attractors $\{ \kappa_{\gamma\gamma'}^\nu \}$ can be bounded, both above and below, by components of entropy production along corresponding paths, viz., \cite{Freitas22,Falasco25}
\eqs{
-\sigma_{\nu \to \gamma} \leq \frac{1}{\Omega} \log \kappa_{\gamma\gamma'}^\nu \leq \sigma_{\gamma \to \nu}, 
}
and these bounds are sharp both in the detailed balance case and to first order in nonconservative forces.  Comparing with Eq.\eqref{kappa} the control theoretic interpretation of $I_{ss}$ then establishes bounds between entropy production and integrated control. In particular the upper bound on $\kappa^\nu_{\gamma\gamma'}$ indicates that for a large entropy drop, strong control is necessary. If local creation of negative entropy is necessary for life \cite{schrodinger}, then control plays an essential role.

Thermodynamic uncertainty relations state that for a current $O$ its variance and mean are related by $\langle O^2 \rangle_c/\langle O\rangle^2 \geq 2/\sigma$ where $\sigma$ is the entropy production in the process, in units of $k_B$ \cite{Barato15,Horowitz20}. This is usually read as saying that precision requires entropy dissipation. Since variances are related to noise, and hence to control, an alternative reading is that minimal entropy dissipation requires strong control \footnote{A similar conclusion was reached in \cite{Still12}, where `control' is replaced by `predictive power,' and quantified with information theory.}. 

This new perpective may help understand the largely unexplained thermodynamic efficiency of biological systems \cite{Kempes17,Wolpert24}. Indeed the notion that in a biochemical system any particular species needs to be maintained at a precise concentration is usually a prejudice; the cell is apparently content to operate with significant copy-number fluctuations \cite{Elowitz02}. What is crucial is that the system continues to play the same functional role. By recasting noise as control, this shifts focus from seeking mechanisms that eliminate all fluctuations, or work in spite of them, to understanding the relationship between control and objective. 

\begin{figure*}[t!]
\centering
\includegraphics[width=\textwidth]{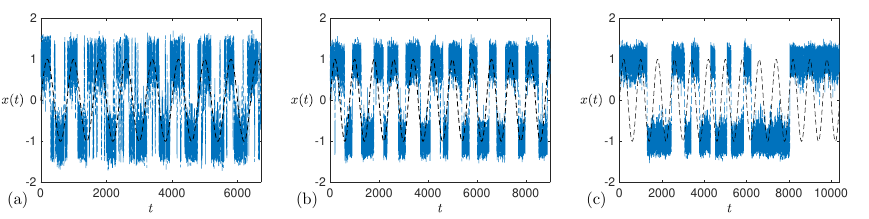}
\caption{ Illustration of double-well system with time-periodic forcing. In all panels, the forcing period is $T=800$. (a) Volume $\Omega=5$ and Kramers time $\tau \approx 54$; (b) Volume $\Omega=9$ and Kramers time $\tau \approx 400$; and (c) Volume $\Omega=12$ and Kramers time $\tau \approx 1800$. Here $a=b=1$ and $A_0=0.1$. Kramers times are defined in absence of periodic forcing, $\tau = \sqrt{2}\pi e^{\Omega/2}$  \cite{Gammaitoni98}. }\label{SR}
\end{figure*}

\section{Unstable fixed points} 

Unstable systems can be stabilized by control. While much of physics is built on harmonic behavior around equilibria, when noise is added and interpreted as control, the restriction to stable equilibria is unnecessarily strict. A growing literature in the ecology \cite{Boettiger18} and systems biology \cite{Eldar10,Hilfinger16} communities indeed finds that noise can act as a stabilizing force. Hitherto these have been given idiosyncratic explanations, if any. The noise-control mapping explored here instead makes it natural. 

{ To see this, consider again the double-well example of Fig. \ref{well1}. The phase portrait in $x-\pi$ space is shown in Fig.\ref{phaseportrait}, for $a=b=1$. In addition to the stable fixed points at $x=\pm \sqrt{a/b}$ and the unstable saddle at $x=0$, all at $\pi=0$, there are additional fixed points visible at nonzero $\pi$. From Eq.\eqref{dblwelleq} we find them at $x = \pm \sqrt{a/3b}, \pi = - 2ax/3$. These are control-stabilized fixed points that do not exist deterministically, reflecting a balance between control and relaxation. When perturbed they perform a cycle, like Sisyphus pushing a boulder up a hill only for it to fall back to the bottom. If we write $x = \pm \sqrt{a/3b} + \delta x, \pi = - 2ax/3 + \delta \pi$ and linearize Eq.\eqref{dblwelleq} around the fixed point, we get 
\eqs{
\p_t \delta x & = \delta \pi \\
\p_t \delta \pi & = - \frac{4}{3} a^2 \delta x 
}
which reduces to $\p_t^2 \delta x = - \frac{4}{3} a^2 \delta x$ and hence oscillations with angular frequency $\omega = a \sqrt{4/3}$. Dynamically, they correspond to most-likely `failed' hops. Evaluating $H$ we get $H/\Omega = -2 a^3/(27b) < 0$ on the fixed point. 

On topological grounds, such control-stabilized fixed points (more generally manifolds) are expected to be generic. Indeed, 
consider a deterministic fixed point $\gamma$ and one of its saddles $\nu$. From $\nu$ down to $\gamma$ there is a manifold along $\piv=0$ giving the relaxation trajectory. The `uphill' trajectory instead is at $\piv \neq 0$, so these two manifolds enclose a volume, with a circulation flux, because their trajectories are oriented in the opposite way. Both of these manifolds have $H=0$. The enclosed volume can be followed in contours of $H$ until it collapses on a manifold of lower dimension, for example a fixed point in 2D. This is the control-stabilized manifold. 


In Appendix 4, we consider more generally the control of unstable linear systems, paying attention to the form of the objective relevant for noisy systems. We show that unstable fixed points found already in the deterministic dynamics are not asymptotically stable under optimal control, and so eventually leave the vicinity of the fixed point. But there are also generally control-stabilized fixed points like in the double-well example. These control-stabilized fixed points capture the essential irreversible dynamics in a way that is impossible at a deterministic fixed point, and may play an important role in finite-time dynamics. 
}

{

\section{Time-dependent forcing}

In the examples above, the drift field was taken as static. As a result, although over a long time scale there will be hops between attractors, each showing goal-directed behavior, there was no imperative to choose one attractor over another. Rich agential phenomena appear as soon as we add time-dependence to the potential. 

Consider a Langevin equation with $F \to \Omega \p_x U(x,t)$ and $B = \Omega^{1/2}$. The Hamiltonian is $H/\Omega =- \pi \p_x U(x,t) + \half \pi^2$. Suppose the potential $U(x,t)$ is slowly-varying in time, and has multiple attractors in space. In the adiabatic limit, we treat $U(x,t)$ as fixed over the duration of a hop, centered at $t_0$. Then the action for a hop from $\gamma$ to $\gamma'$ is
\eq{
S/\Omega & = \int_{\gamma}^{\gamma'} \frac{dx}{\p_t x} \left[ \pi \p_t x - H' \right] = \int_{\gamma}^{\gamma'} dx \left[ \pi - \frac{H'}{-\p_x U + \pi} \right] \notag \\
& = \half \int_{\gamma}^{\gamma'} dx \;\frac{\pi^2}{\pi - \p_x U} ,
}
with $\pi(x) = \p_x U \pm \sqrt{ (\p_x U)^2 + 2 H'}$, and the potential evaluated at $t=t_0$. The two branches correspond to right-going and left-going hops. For simplicity consider slow hops $H=0$, which have minimal action. Then 
\eqs{
S/\Omega \to 2(U(x_\nu,t_0) - U(x_\gamma,t_0)) 
}
where $\nu$ is the saddle. The action reduces to twice the barrier height, and minimization of action over $t_0$ will then pick out the minimal barrier. 

For example, if the potential is the sum of a double well and a time-periodic rocking ramp, $U(x,t) = -\half a x^2+ \frac{1}{4} b x^4 - x A_0 \sin(\omega t)$, then a hop starting in one well at $t=0$ will occur when $\omega t_0 = \pi/2$, which is the first trough of the potential. This mechanism for phase-locking is the kernel of stochastic resonance \cite{Benzi81,Gammaitoni98,McDonnell09}: random hops that would occur erratically without temporal forcing become entrained by a weak periodic signal. By the noise-control mapping, {\it the system chooses the optimal time to hop}. Functional agential behavior appears naturally. 

A second feature of stochastic resonance is that the effect is largest at an intermediate noise level. This follows straightforwardly: we can stitch together an oscillation of the noisy system with two hops, one forward and one backward. Since the optimal hops will occur at the peak and trough of the potential, the total time is one full period, $T = 2\pi/\omega$, and each hop should occur in a time $T/2$. But the most likely time-to-hop, $\tau$, is given by the Kramers expression $\tau \propto e^{S}$ \cite{Gammaitoni98}, and depends on the noise level. If $\tau > T/2$, then hops will be less frequent than optimal, while if $\tau < T/2$, then they are more frequent than optimal. At $\tau=T/2$ the resonance will be largest. 

This phenomenon is illustrated in Fig.\ref{SR} for three different noise levels at a fixed driving period. Although the Kramers times vary over a considerable range, both above and below half the driving period, there is strong entrainment of the periodic signal. Entrainment is largest when $\tau \approx T/2$ as expected; when noise is weaker, hops are missed, while when noise is stronger, there are extra hops.

\begin{figure}[t!]
\centering
\includegraphics[width=0.7\columnwidth]{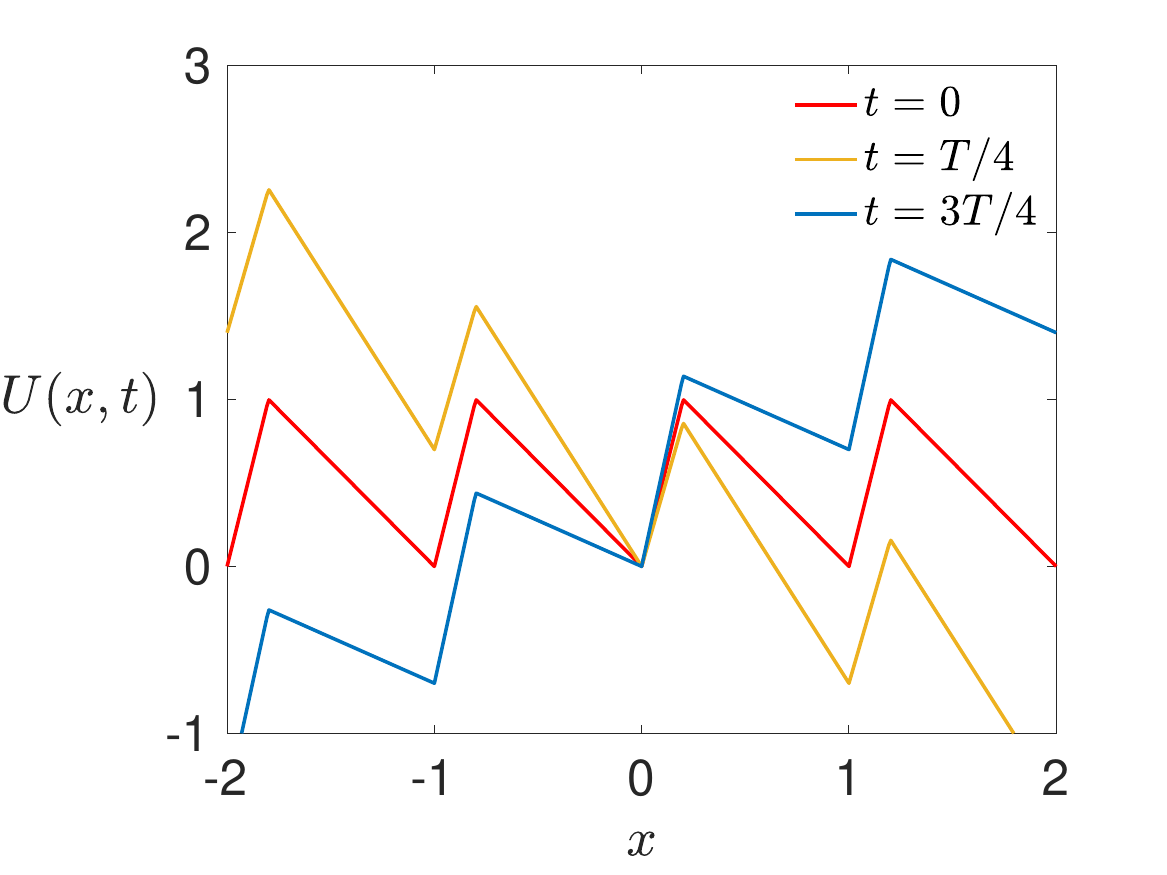}
\caption{ Illustration of a sawtooth potential periodically rocked in time at indicated times $t$, as a fraction of the period $T$. The figure is shown with $a=0.2$ and $A_0=0.7$. The periodic modulation of the barrier height is evident at $x=0.2$. }\label{sawtooth}
\end{figure}

As a second example, let us show that one additional feature opens up another rich set of connections to agential behavior. Consider a time-periodic potential but now let $U$ have a spatially-periodic and {\it asymmetric} form, like a sawtooth. To be explicit we will consider a rocking potential $U(x,t) = u(x) - x A_0 \sin (\omega t)$ where 
\eqs{
u(x) = \Delta u \begin{cases}  (x/a) &  0<x<a \\
(1-x)/(1-a) & a<x<1 
\end{cases}
}
and $u(x+1)=u(x)$. For $0<a<1/2$ the sawtooth is steeper for rightgoing hops, as pictured in Fig.\ref{sawtooth}. Minimizing the action for a hop starting at $t=0$ again leads to $\omega t_0=\pi/2$, but now the barrier height is asymmetric in left and right hops: $S_+/\Omega =  2 \Delta u - 2 a A_0$ for a rightward hop, and $S_-/\Omega =  2 \Delta u - 2 (1-a) A_0$ for a leftward hop. When the potential is not symmetric $a \neq 1/2$, there is net movement at a rate proportional to $\approx e^{-2\Omega \Delta u} [e^{2 \Omega a A_0} - e^{2 \Omega (1-a) A_0}]$. {\it Directional motion is rectified from thermal fluctuations.} Moreover, like stochastic resonance, the effect is non-monotonic with the noise strength \cite{Astumian94}. This is the underlying mechanism behind a Brownian ratchet in the weak-noise regime \cite{Magnasco93,Astumian94}, a model of a molecular motor. By the noise-control mapping, an agential interpretation of the results is justified: the system chooses the optimal times to hop, and in so doing achieves net movement. 

These examples can be considered as building blocks of rich agential behavior. Indeed, stochastic resonance occurs not only with strictly periodic forcing, but also aperiodic forcing, so long as the timescale of the forcing is much slower than the internal relaxation timescales of the system \cite{Collins95,Collins96}. For example, \cite{Collins96} considers a double-well subject to an arbitrary modulation of the barrier. 

In the general situation of a stochastic system with multiple attractors subject to external time-dependent forcing, we can expect some degree of entrainment of the forcing. In the weak-noise regime, the system chooses optimal times to hop such that entrainment is maximal. The external forcing can even be stochastic: in this case, the forcing acts as exogenous control, and cooperates with endogenous control. 

\section{Discussion}

\subsection{Two analogies}

The main result of this work is that agency is part of Nature. Since this appears at odds with our conception of a causal Nature, we try to illuminate the paradoxical aspects by way of analogy.

The first analogy is with the emergence of dissipative behavior in thermodynamics. If, for example, a gas is modelled by Newton's laws with conservative interactions, then it is time-reversal symmetric at the microscopic level. However, the emergent macroscopic thermodynamic behavior is dissipative, and breaks time-reversal invariance. There is no contradiction, because in thermodynamics the time-reversal symmetry is broken by the initial conditions, that is we typically begin with a state out of thermal equilibrium. Moreover, the microscopic description is more complex than the macroscopic description precisely because it retains memory of the fundamentally conservative dynamics.

Similarly, in the present work we begin with a dissipative system with causal dynamics, but on long time scales there is emergent agential and acausal behavior. There is no contradiction, because one cannot precisely predict {\it when} the agential hops will occur. Moreover, the initial dissipative description is more complicated than the emergent description precisely because it retains causality. 

The second analogy is with the double-slit experiment of quantum mechanics. In the classic experiment, and assuming the Copenhagen interpretation of quantum mechanics, it is {\it forbidden} to ask which slit an individual electron went through. Similarly, in the present work, the hops are agential and acausal but it is forbidden to ask precisely when they occur. 

\subsection{Optimization principles}

Inspired by the power of extremal principles in thermodynamics, there has been a long history of attempts to find extremal principles in nonequilibrium statistical physics, and to find them in biology in particular. We can divide these into two types: 

(i) Maximum entropy methods \cite{Bialek12,Mora10,Schneidman06} can be viewed as attempts to quantitatively capture the probability distribution of a system, given the values of some observables, without unnecessary modelling assumptions. Success of a maximum entropy method, measured by capture of patterns not used as input to the method (such as higher-order correlations), is then interpreted as typicality of the data given the observables used as input. The method is agnostic as to the underlying mechanism through which Nature achieves the probability distribution. For example, the maximum entropy method applied to a conservative system with energy used as a constrained observable, reproduces the canonical ensemble, but is silent on the ergodicity-generating (scrambling) dynamics that, from a reductionist perspective, underlie its validity. 

(ii) Alternatively, it has been proposed that biological systems may be optimal with respect to some functional requirement: precise examples include minimal dissipation cost \cite{Han08}, optimal information flow \cite{Tkacik16,Tkacik25}, minimal protein production cost \cite{Hu20}, and optimal growth (\cite{Orth10,Dourado20} and references therein). In these cases no explicit mechanism from mechanics or thermodynamics is available to justify optimality; it is implicit or explicit that natural selection is the underlying driver. Regardless of any empirical success of optimality, this puts an enormous explanatory burden on natural selection, which is difficult if not impossible to remove. 

In a weak-noise system, transition probabilities are governed by optimal control problems. Each transition minimizes action, which for slow transitions becomes integrated control (c.f. Eq.\eqref{kappa}). Can optimality of transition probabilities shed light on global optimality? 

An example given earlier illustrates the difficulty: in a double-well system, the infinite-time transition probability from $-1$ to $+1$, $\PP(+1|-1, t \to \infty) = 1/(1 + \kappa_{1,-1}/\kappa_{-1,1})$ in terms of the forward and backward transition rates. Each of these rates is governed by an optimal control problem, but maximizing the transition rates does not in itself maximize or minimize $\PP(+1|-1, t \to \infty)$. And in systems with many attractors, the infinite-time transition probabilities are nearly intractable functions of the individual rates, so this relationship becomes even more intricate. 

However, we can find observables that would be maximized when all $\{\kappa_{\gamma \gamma'} \}$ are maximized, where $\gamma,\gamma'$ label attractors. Indeed if
\eq{
\Theta = \sum_{\gamma,\gamma'} f_{\gamma\gamma'}(\kappa_{\gamma\gamma'})
}
where all $f_{\gamma\gamma'}$ are increasing, then clearly $\Theta$ is maximized when the $\{\kappa_{\gamma \gamma'} \}$ are. Do any physical observables take this form? For example, when $f_{\gamma\gamma'}(x)=f(x)=x\log x$ then $-\Theta$ takes the form of an entropy functional, but this has no accepted physical meaning, since the $\{\kappa_{\gamma \gamma'} \}$ are rates, not probabilities. 

In general, the relationship of a global observable to the optimization problems fixing the $\{\kappa_{\gamma \gamma'} \}$ is that of {\it decentralized control} \cite{Siljak11}, since these optimization problems are independent, and local. In future work it would be useful to see when global physical observables can be nontrivially extremized by maximization of the $\{\kappa_{\gamma \gamma'} \}$. 

}

\section{Conclusion} 

{ 

We now return to the two conceptual problems raised in the Introduction. The first was that path integral formulations of dissipative stochastic dynamics represent arbitrary observables in terms of trajectories that are real-valued and in physical space but involve an acausal field $\pi$ whose interpretation is obscure. The resolution is that acausality is real: the $\pi$ field represents the control applied by noise. Concretely, Nature is always sampling control strategies. Agency is part of Nature, in the precise sense of the noise-control mapping. This also then solves the second problem, of agency-body duality, since the agential description is obtained as a bijective mapping from a mechano-stochastic one, so that there is no fundamental conflict between them. 

We emphasize that the origin of agency in Nature is not merely a problem of interpretation: goal-directed behavior necessarily involves acausal trajectories, and how these are generated from fundamentally causal dynamics is a nontrivial physical problem. Here we have shown how and why they arise naturally from noise-induced hops in any multi-attractor system. When hops occur, they will appear to observers to be goal-directed over the timescale of the hop.

}


Finally, although the cybernetic approach \cite{Wiener19,Ashby56,Ashby58} pioneered the use of control theory in biology, it waned as it became divorced from the spectactular successes of reductionist molecular biology. The noise-control mapping elucidated here grounds cybernetic ideas in a precise connection, valid at any scale. It also behaves well under coarse graining, because the long-time dynamics of a CRN is itself a jump process over attractors \cite{Vellela09,Smith20,Falasco25}. What remains is to understand the generic forms of the objective in realistic systems, and especially how these relate to dynamics (point (ii) in \cite{Dehghani24}). Indeed the noise-control mapping is expected to be most fruitful in the cases that are difficult to describe in any language, for example when the dynamics visits many metastable states, and is highly sensitive to external forcing. 






\begin{acknowledgments}

I am grateful to AI Brown for comments on the manuscript. EDG is supported by NSERC Discovery Grant RGPIN-2020-04762. 
\end{acknowledgments}

\begin{figure*}[t!]
\centering
\includegraphics[width=\textwidth]{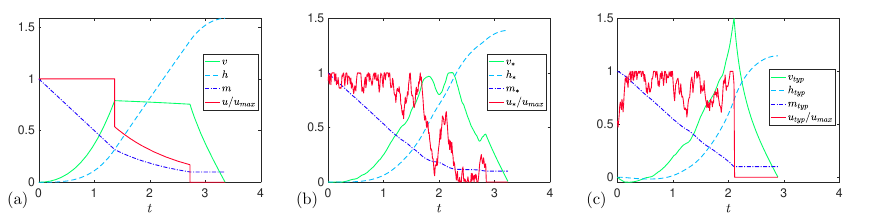}
\caption{ Control problems can be solved by stochastic systems. (a) Optimal control for Goddard problem, with 3 stages plainly visible; (b) Best trajectory (highest final height) in noisy Goddard system over a sample ensemble of 1000 trajectories; (c) Typical trajectory at $\beta = 5$. These results take units with $g=h_0=m_0=1$ and parameters $u_{max}=1, c=\gamma=1/2, m_1 = 0.1$. }\label{Goddard}
\end{figure*}

{\bf Appendices. } \\ 

Here we describe the Goddard rocket problem and its solution by a Langevin process (Appendix 1); the analytical approach to obtaining fluctuations around instantons, and their control-theoretic interpretation (Appendix 2); the use of the PMP to rederive the Hamiltonian equations for a CRN (Appendix 3); and the analysis and interpretation of unstable fixed points (Appendix 4). \\

{\bf Appendix 1. Goddard rocket problem: }\\

Consider a rocket with vertical velocity $v(t)$ at height $h(t)$, with total mass $m(t)$ (including both fuel and chassis). They evolve according to Newton's laws
\begin{subequations} \label{God} \eq{ 
\p_t v & = \frac{1}{m} \left[ u  - D(v,h) \right] - g \\
\p_t h & = v \\
\p_t m & = -\gamma u ,
} \end{subequations}
where $u(t)$ is the thrust control, $D(v,h)$ is the drag, and $\gamma$ is an efficiency coefficent. We start at sea level $v(0)=h(0)=0$ and with mass $m(0)=m_0$. The Goddard problem is to maximize the height obtained, $h(t_f)$ where the final time is when the fuel runs out $m(t_f)=m_1$ and the velocity vanishes $v(t_f)=0$, where it is understood that $m_1$ is the mass of the chassis only. The thrust is subject to $0 \leq u \leq u_{max}$. 

The solution to this optimal control problem is highly nontrivial \cite{Glad18}. Indeed, although naively it might seem optimal to burn fuel as quickly as possible, to avoid carrying it unnecessarily to higher altitudes, when drag is factored in this is not necessarily true: it may be worth it to carry some fuel higher, where air is thinner and drag is reduced (or simply to burn it at a slower speed). It turns out that the optimal trajectory has generally 3 stages: a first `full throttle' stage in which $u=u_{max}$; a second stage in which the control is nontrivial; and a final stage in which fuel is spent and the rocket decelerates to its maximum height. 

Here we want to show that this solution can be approximated by a Langevin process, which we call the noisy Goddard system. We simply take Eq.\eqref{God} and model $u$ as noise. We consider it to be continuous with uncorrelated increments $du \in [-1,1] \times \sqrt{dt}$ where in the numerics we have $dt=0.01$. We impose $0 \leq u(t) \leq u_{max}$. To compare with the optimal control, we turn the control off once $m(t)=m_1$. 

In the numerics we take units in which $g=m_0=1$, and we fix a drag law $D(v,h) = c v^2 e^{-h/h_0}$. We fix a length by $h_0=1$. In this problem there is no natural way to scale the noise that would still admit solutions in the weak-noise regime: for example if $u_{max}$ is too small then the rocket cannot gain altitude. In the numerics we sample paths with $e^{\beta h(t_f)}$ where $\beta$ can be adjusted. (The optimal control is of course independent of $\beta$.) 

Fig.\ref{Goddard} shows the results. At left is shown the optimal control with its 3 stages; at center is the best trajectory in a run of 1000 noisy trajectories (this result is typical over ensembles of this size); and at right is a median trajectory closest to the ensemble at $\beta = 5$. These 3 trajectories give $h(t_f) = 1.58, 1.39,$ and $1.15$, respectively. We see that the optimal control is well approximated by a noisy trajectory and comes within 15\% of the maximum height. These results are typical, and illustrate that this control problem can be solved -- within approximation, at a finite number of trajectories, and finite $\beta$ -- by a noisy system. 

\begin{figure}[t!]
\centering
\includegraphics[width=\columnwidth]{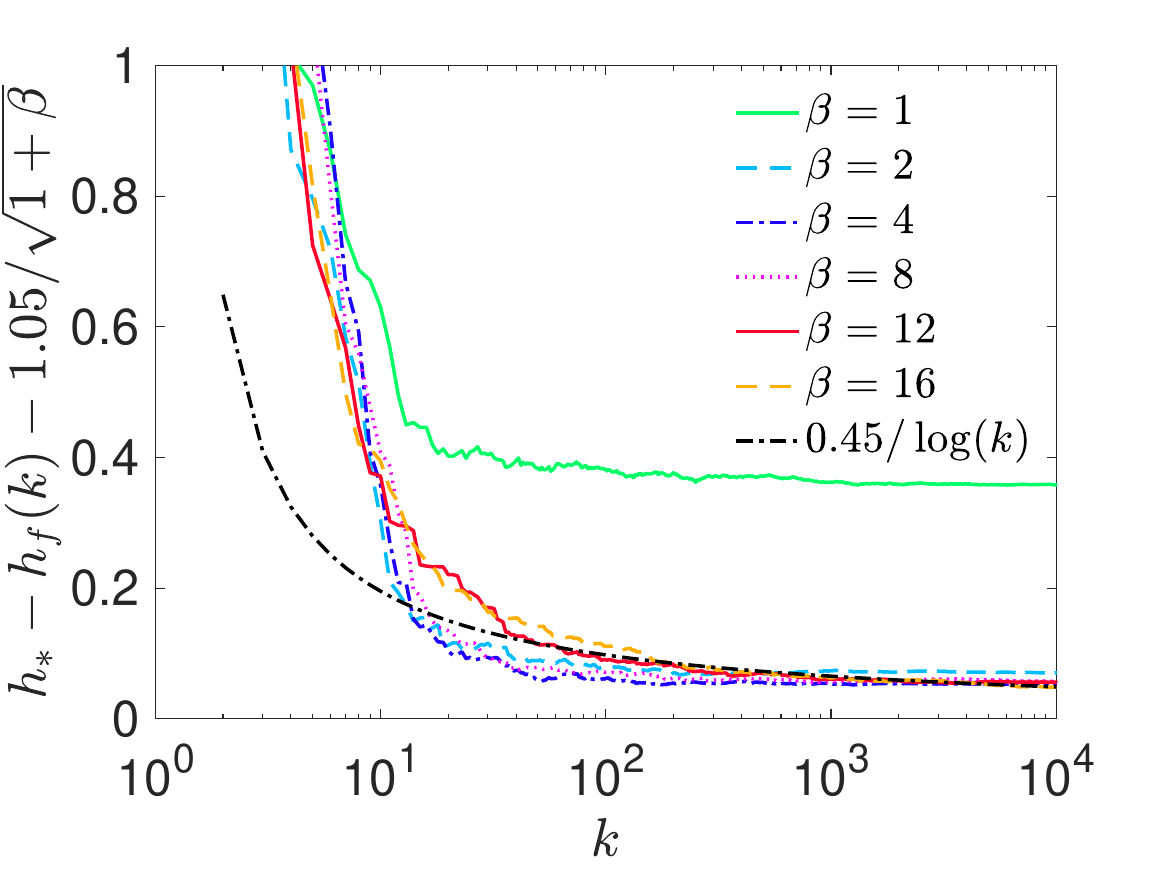}
\caption{ Approach to optimal control in the noisy Goddard system, over typical ensemble of size $k$, at indicated $\beta$. We have subtracted off an asymptotic value $1.05/\sqrt{1+\beta}$. These results take units with $g=h_0=m_0=1$ and parameters $u_{max}=1, c=\gamma=1/2, m_1 = 0.1$. }\label{Goddard2}
\end{figure}

{ The approach to the optimal control is quite slow. Over an ensemble of trajectories, define the $j$th height reached $h_j$ and the running average $h_f(k)$ for the trajectories $1,\ldots,k$, i.e.
\eqs{
h_f(k) = \frac{\sum_{j=1}^k h_j e^{\beta h_j} }{\sum_{j=1}^k e^{\beta h_j} }.
}
Then define a typical value $dh$ of $h_* - h_f(k)$ by averaging over 50 different shufflings of the trajectories, which represents the typical performance at a given $k$. We measured $dh$ for $\beta = 1,2,4,8,12,16$ over an ensemble of 10000 trajectories. At finite $\beta$, $dh$ will converge as $k \to \infty$ to a value that we find $\approx 1.05/\sqrt{1+\beta}$. As shown in Fig.\ref{Goddard2}, for $\beta > 1$ the approach is well-approximated by $0.45/\log k$, so that finally
\eqs{
h_f(k) \approx h_* - \frac{1.05}{\sqrt{1+\beta}} - \frac{0.45}{\log k}.
}
For any $\beta$ we need to take $k$ very large to approach the asymptotic value. \\
}

{\bf Appendix 2. Gaussian fluctuations around instantons: }

Consider, either for Langevin dynamics or Markov jump processes,
\eqs{
S(\nv,\piv) = \int dt \; \left[ \piv \cdot \p_t \nv - H(\nv,\piv) \right]
}
where $H(\nv,0)=0$ and we suppress boundary conditions $\nv(0)=\nv_0, \nv(t_f)=\nv_f$. 

For Langevin dynamics $H$ takes the form
\eqs{
H(\nv,\piv) = -\piv \cdot \vec F(\nv) + \half \piv \cdot B \cdot B^T \cdot \piv \qquad \text{Langevin}
}
while for CRNs each reaction $\alpha$ of the form $\sum p_{\alpha j} X_j \to \sum q_{\alpha j} X_j$ ($X_j$ denote the species) contributes
\eqs{
H_\alpha(\nv,\piv) = (e^{\piv \cdot (\vec q_\alpha-\vec p_\alpha)}-1) \tilde k_\alpha \prod_i n_i^{p_{\alpha i}} \qquad \text{CRN}
}
where $\tilde k_\alpha$ is a rescaled reaction rate (see \cite{De-Giuli22b}). These specific forms will not be needed in this section. 

{ We adopt a vector notation such that all contractions are explicitly indicated, e.g. $\vec{a} \vec{b}$ is a matrix with entries $a_j b_k$; and $\vec a \vec b : \hat A$ is $\sum_{j,k} a_j b_k A_{jk}$ if $\hat A$ is a matrix of the appropriate size. }

We set $\nv = \nv_* + \nv'$ and $\piv = \piv_* + \piv'$ and look at the regime of small $\nv', \piv'$, which is relevant in the macroscopic limit. To quadratic order in $\nv',\piv'$ we obtain
\eqs{
S(\nv,\piv) & = S_* + \int dt \; \left[ \piv_* \cdot \p_t \nv' + \piv' \cdot \p_t \nv_* + \piv' \cdot \p_t \nv' \right.  \notag\\ 
& \left. - \nv' \cdot \nabla_n H  - \piv' \cdot \nabla_\pi H - \nv' \piv' : \nabla_n \nabla_\pi H \right.  \notag\\ 
& \left. - \half \nv' \nv' : \nabla_n \nabla_n H  - \half \piv' \piv' : \nabla_\pi \nabla_\pi H  \right] + \ldots
}
where $S_*=S(\nv_*,\piv_*)$ and the derivatives of $H$ are evaluated on $(\nv_*,\piv_*)$. Choosing the latter to satisfy the instanton equations Eq.\eqref{ham} this is reduced to
\eqs{
S & \to S_* + \int dt \; \left[  \piv' \cdot \p_t \nv' - \piv' \nv' : A \right. \\
& \left. \qquad - \half \nv' \nv' : K  - \half \piv' \piv' : C \right] \\
& \to S_* + \int dt \; \piv' \cdot \left[ \p_t \nv' - \nv' \cdot A - \xiv \cdot C \right] \\
& \qquad + \half\int dt \; \left[ - \nv' \nv' : K  + \xiv \xiv : C \right] 
}
where in the last step we performed a Hubbard-Stratonovich transformation to introduce $\xiv$. Here
\eqs{
A & = \nabla_\pi \nabla_n H\\ 
K & = \nabla_n \nabla_n H\\
C & = \nabla_\pi \nabla_\pi H .
}
If we integrate out $\piv'$ then we obtain a new Langevin equation that is linear in $\nv'$ but generally non-autonomous, viz.,
\eq{ \label{ham2}
\p_t \nv' - A \cdot \nv' = C \cdot \xiv 
}
Moreover the objective function now has two terms
\eq{ \label{S2}
S_2 = \int dt \; \left[ - \half \nv' \nv' : K  + \half \xiv \xiv : C\right]
}
Now since $H(\nv,0)=0$, we have $K=\nabla_n \nabla_n H=0$ on deterministic trajectories $\piv_*=0$: it is only present under active control.

The remaining problem is to integrate out $(\nv',\xiv)$ (or equivalently $(\nv',\piv')$). If the boundary conditions on $\nv'$ are null, because they have already been accounted for in $\nv_*$, then the result will only contribute the term in $S_2$ quadratic in $\nv'$, and a (functional) determinant, which can be evaluated \cite{Gelfand60,Dunne08}. 


For simplicity consider the case when $(\nv_*,\piv_*)$ is a fixed point, so that $A,K,C$ are all constant matrices. Then the result is \cite{Schorlepp21}
\eq{ \label{Pnf2} 
\PP(\nv_f | \nv_0) \propto e^{-S_*} e^{\hhalf \int Q : K} \left[ \det Q(t_f) \right]^{-1/2}
}
where $Q(t)$ satisfies 
\eq{ \label{Q2}
\p_t Q = Q \cdot A^T + A \cdot Q + C + Q \cdot K \cdot Q
}
with $Q(0)=0$. Eq.\eqref{Pnf2} also holds in more general conditions when the dynamics is nonlinear and can include multiplicative noise. In the latter case some renormalization of the determinant is necessary \cite{Schorlepp25}. 

To understand the physical meaning of $Q$, we consider the Hamiltonian form of Eqs.(\ref{ham2},\ref{S2}), viz.,
\eqs{
\p_t \nv' & = A \cdot \nv' + C \cdot C^T \cdot \piv \\
\p_t \piv & = - A^T \cdot \piv .
}
Construct a basis set of solutions satisfying $\nv'^{(k)}(0)=0, \pi^{(k)}_j(0)=\delta_{jk}$. These correspond to all the directions that can be reached from $\nv'(0)=0$. These can be placed as columns into matrices $\delta N, \delta \Pi$. Then the matrix
\eq{ \label{Qp}
Q' = \delta N \cdot (\delta \Pi)^{-1}
}
satisfies the same equation Eq.\eqref{Q2}. Since $\delta N(0)=0$, we get also that $Q'(0)=0$, so $Q'=Q$.

From Eq.\eqref{Qp} we can write $\delta \Pi = (Q')^{-1} \cdot \delta N$, meaning $\piv^{(k)} = (Q')^{-1} \cdot \nv^{(k)}$. In other words, these solutions take a feedback form where $(Q')^{-1}$ is the feedback matrix giving the controls in terms of the state fluctuation. 

This argument extends straightforwardly to any situation for which Eq.\eqref{Pnf2} holds, including cases where the instanton $\nv_*$ is time-dependent. \\


%
%

{\bf Appendix 3: Pontryagin Maximum Principle: } \\

The Pontryagin maximum principle (PMP) states that, for Eqs.(\ref{eq1},\ref{eq2}) and general $L$, the optimal control must maximize the Pontryagin Hamiltonian
\eqs{
H_P(\nv,\xiv,\pv,t) = \pv\cdot [B(\nv)\cdot \xiv - \vec F(\nv)] - L[\nv,\xiv]
}
pointwise in $\xiv$, where $\pv$ solves the costate equations
$\p_t \pv = - \nabla_n H_P$. In the Langevin case $L=\half \xiv^2$ the maximization of $H_P$ leads to $\xiv = \pv \cdot B(\nv)$ and then $H_P \to \half \pv \cdot B \cdot B^T \cdot \pv + \pv \cdot \vec F = H$ when we identify $\pv = \piv$. As expected we recover Eq.\eqref{ham} but without path integral manipulations.

For Markov jump processes we consider the PMP with objective $\int L$ with $L(\nv,\piv) = \piv \cdot \nabla_\pi H - H$ and field equation $\p_t \nv = \nabla_\pi H$. We have
\eqs{
H_P(\nv,\piv,\pv,t) & = \pv \cdot \nabla_\pi H - L(\nv,\piv) \\
& = (\pv-\piv) \cdot \nabla_\pi H + H(\nv,\piv) 
}
and the costate equation is
\eqs{
\p_t \pv & = - \nabla_n H_P \\
& = (\piv-\pv) \cdot \nabla_\pi \nabla_n H - \nabla_n H .
}
Extremizing $H_P$ over $\piv$ we get
\eqs{
0 = \nabla_\pi H_P = (\pv-\piv) \cdot \nabla_\pi \nabla_\pi H , 
}
solved by $\pv = \piv$. Then the costate equation reduces to $\p_t \piv = - \nabla_n H$, which is the correct equation. Moreover the objective becomes $\int L \to \int [ \piv \cdot (\p_t \nv) - H] = S$ as required. \\

{\bf Appendix 4: Unstable modes: } \\

Consider the linear system Eq.\eqref{ham2} with objective Eq.\eqref{S2}, where the matrix $A$ may have unstable modes. We let the boundary conditions be $\nv'(0)=\nv_0, \nv'(t_f)=\nv_f$. We are assuming that $A,C,$ and $K$ are time-independent, which will hold if we apply this formalism to the linear dynamics around fixed points. Importantly, these may include non-classical fixed points where $\piv_* \neq 0$, as discussed later. In the language of control theory this is a linear-quadratic regulator (LQR). (One can also drop the final boundary condition and replace it with a cost on the final state). 

Consider the feedback form
\eq{ \label{feed}
\xiv = -R \cdot \nv' 
} 
{ defining a feedback matrix $R$, }
so that the so-called closed-loop dynamics for $\nv'$ becomes
\eqs{
\p_t \nv' = \underbrace{(A - C \cdot R)}_{M} \cdot \nv' .
}
This closed-loop dynamics will be asymptotically stable if $M$ is Hurwitz, i.e. all its eigenvalues have negative real part. It is known from control theory that if the system is `controllable' (i.e. if any state can be reached by some control, which is expected to be the usual case in noisy systems), then we can define $R = C^T \cdot W^{-1}$ where
\eqs{
W(t) = \int_0^{t} ds \; e^{-sA}\cdot C \cdot C^T \cdot e^{-s A^T} ,
}
and $M$ will be Hurwitz \footnote{An analogous but less explicit Theorem also holds in the nonlinear case. (\cite{Sontag13} Thm.19)}. Note that 
\eqs{
A \cdot W(t) & = -\int_0^{t} ds \; \p_s [ e^{-sA} ] \cdot C \cdot C^T \cdot e^{-s A^T} \\
& = - e^{-tA}\cdot C \cdot C^T \cdot e^{-tA^T} + C \cdot C^T \notag\\
& \qquad + \int_0^{t} ds \;  e^{-sA} \cdot C \cdot C^T \cdot \p_s [  e^{-s A^T} ] \\
& = - \p_t W(t) + C \cdot C^T - W(t) \cdot A^T ,
}
i.e.
\eq{ \label{W}
\p_t W(t) = A \cdot W(t) + W(t) \cdot A^T + C \cdot C^T
}
To illustrate, we focus on the case when $A$, $C$, and $K$ are the same size, and simultaneously diagonalizable, so that modes are independent. Then we write
\renewcommand{\l}{\lambda}
\eqs{
A & = \sum_\l \l |\l\rangle \langle \l | \\
C & = \sum_\l c_\l |\l\rangle \langle \l | \\
K & = \sum_\l k_\l |\l\rangle \langle \l | \\
W & = \sum_\l w_\l(t) |\l\rangle \langle \l | 
}
where we adopt a bra-ket notation. We assume that $A$ has no zero modes; the limit $\l \to 0$ can be taken later if necessary.
For each mode we have
\eqs{
w_\l(t) = c_\l^2 \int_0^{t} dt e^{-2t\l} = \frac{c_\l^2}{2\l} \left[ 1 - e^{-2t \l} \right]
}
and we find that the eigenvalues of $M$ are 
\eqs{
m_\l(t) & = \l - c_\l^2 w_\l(t)^{-1} \\ 
& = \l - \frac{2 \l}{1 - e^{-2t\l} } \\
&  \xrightarrow[t \to \infty]{} \begin{cases} -\l & \mbox{Re}[\l]>0 \\ +\l & \mbox{Re}[\l]<0 \end{cases}
}
so that the system is stabilized. However, this form of regulator is not necessarily the optimal control, because it ignores the objective, and will not generically fit the boundary conditions.

It turns out that the optimal control still has the form Eq.\eqref{feed}, but the matrix $R$ satisfies
\eqs{
\p_t R = - A^T \cdot R - R \cdot A + R \cdot C \cdot R + K 
}
If we define 
\eqs{
Q = -R^{-1}
}
and use the identity $\p_{t} R^{-1} = - R^{-1} \cdot \p_{t} R \cdot R^{-1}$ then we obtain
\eqs{
\p_t Q & = R^{-1} \cdot \p_{t} R \cdot R^{-1} \\
& = R^{-1} \cdot \left[ - A^T \cdot R - R \cdot A + R \cdot C \cdot R + K \right]\cdot R^{-1} \\
& = Q \cdot A^T + A \cdot Q + C + Q \cdot K \cdot Q
}
which is exactly Eq.\eqref{Q2} above. We note the similarity with Eq.\eqref{W}, but also the differences, namely the absence there of a quadratic term, whose role will be explained below.

\begin{figure*}[t!]
\centering
\includegraphics[width=\textwidth]{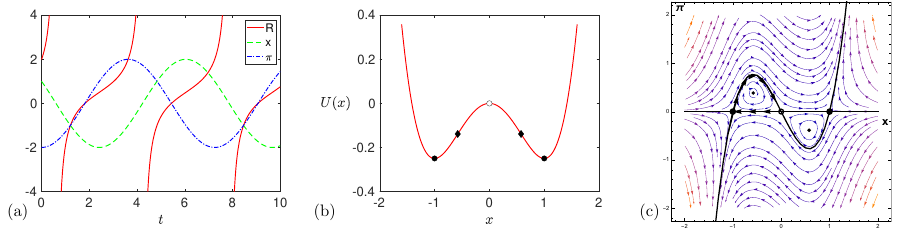}
\caption{ (a) Illustration of cycle around control-stabilized unstable fixed point. $R$ is defined from $\pi(t) = - R(t) x(t)$. (b) Double well potential, with deterministic (circles) and control-stabilized (diamonds) fixed points indicated. (c) Phase portrait of double well oscillator. The $H=0$ manifold is shown in bold. This includes both deterministic trajectories at $\pi=0$ and the uphill instantons. The (uphill) instanton from $x=-1$ to $x=0$, and the relaxation trajectory from $x=0$ back to $x=-1$ are indicated with arrows. The cycles around the control-stabilized fixed points are clear. }\label{well}
\end{figure*}

Writing
\eqs{
Q & = \sum_\l q_\l(t) |\l\rangle \langle \l | \\
\nv' & = \sum_\l x_\l(t) |\l\rangle ,
}
for each mode we have
\eqs{
\p_t q_\l = 2 \l q_\l + c_\l + k_\l q_\l^2
}
with solution
\eqs{
q_\l(t) & = \begin{cases} -\frac{c_\l}{2 \l} + c_{1,\l} e^{2 \l t} & k_\l =0 \\
 \frac{-\l}{k_\l} + \frac{\sqrt{c_\l k_\l-\l^2} }{k_\l} \tan \left((t-t_{0,\l}) \sqrt{c_\l k_\l-\l^2} \right) & k_\l \neq 0
 \end{cases}
}
where $c_{1,\l}$ and $t_{0,\l}$ are fixed by boundary conditions.

The cases $k_\l=0$ and $k_\l \neq 0$ are fundamentally different. 

{\it Case (i):  $k_\l=0$ }: For $k_\l=0$ we find
\eqs{
m_\l(t) & = \lambda + c_\l q_\l(t)^{-1} \\
& = \lambda + c_\l \frac{1}{ -\frac{c_\l}{2 \l} + c_{1,\l} e^{2 \l t} } 
}
Write $c_{1,\l} = -\frac{c_\l}{2\l} e^{-2\l t_f} c'_\l$ so that
\eqs{
m_\l(t) & = \l \left[ 1 - \frac{2}{ 1 + c'_{\l} e^{2 \l (t-t_f)} } \right]
}
Now we have
\eqs{
\p_t [\log x_\l(t)] = m_\l(t) 
}
which is integrated to obtain
\eqs{
x_\l(t) = x_\l(0) e^{\l t} \l \left( c'_\l + e^{-2 \l (t-t_f)} \right)
}
We are primarily interested in the stability problem $\nv'(t_f)=0$, i.e. $x_\l(t_f)=0$. We see that if $c'_\l=-1$ then we will have $x_\l(t_f)=0$. This is so despite the fact that
\eqs{
m_\l(t) & \xrightarrow[t \to \infty]{} \begin{cases} +\l & \text{Re}[\l]>0 \\ -\l & \text{Re}[\l]<0 \end{cases}
}
which is asymptotically {\it unstable}! This means that for a given horizon $t_f$ we can always control the system back to the fixed point, but if the control protocol remains on, then eventually the system will leave the fixed point. To control unstable fixed points with $k_\l=0$ we then need to be eternally vigilant. Contrast this with the second case

{\it Case (ii):  $k_\l\neq 0$ }: Now we find
\begin{widetext}
\eqs{
m_\l(t) = \l + \begin{cases} \frac{c_\l k_\l}{-\l+ \sqrt{c_\l k_\l-\l^2}  \tan \left((t-t_{0,\l}) \sqrt{c_\l k_\l-\l^2} \right) } & \text{Re}[c_\l k_\l-\l^2]>0 \\
\frac{c_\l k_\l}{-\l- \sqrt{\l^2-c_\l k_\l}  \tanh \left((t-t_{0,\l}) \sqrt{\l^2-c_\l k_\l} \right) } & \text{R}e[c_\l k_\l-\l^2]<0 \end{cases}
}
\end{widetext}
whose behavior depends on the sign of Re$[c_\l k_\l-\l^2]$. 

If Re$[c_\l k_\l-\l^2]<0$, then $m_\l(t) \to \sqrt{\l^2-c_\l k_\l}$ as $t \to \infty$, so it is asymptotically unstable, and all the caveats of the first case apply.

Instead if Re$[c_\l k_\l-\l^2]>0$, then the system oscillates in a highly nonlinear way. Moreover, it is insensitive to the boundary conditions so the control protocol can remain on. This regime is quite remarkable in that $m_\l(t)$ has infinitely many singularities. An example is shown in Fig.\ref{well}a for $\l=1/2,c_\l=1,k_\l=2$. The regulator is out of phase with the state, so the feedback has singularities.


To see these ideas in action, we consider a double well potential $U(x) = -\half a x^2 + \frac{1}{4} b x^4$, pictured in Fig.\ref{well}b. The Hamiltonian is
\eqs{
H(x,\pi) = \pi (ax-bx^3) + \half \pi^2
}
and the instanton equations are
\eqs{
\p_t x & = a x - bx^3 + \pi \\
\p_t \pi & = - \pi (a - 3b x^2)
}
The phase portrait is shown in Fig.\ref{well}c.

Looking for fixed points we find five: the two stable minima, $x = \pm \sqrt{a/b}$, with $\pi=0$; the unstable saddle $x=0$ with $\pi=0$; and two non-classical fixed points $x = \pm \sqrt{a/3b}$ with $\pi = -2ax/3$. Now as mentioned above, due to the general property $H(\nv,0)=0$, we get that $K=0$ on all deterministic fixed points, where $\pi=0$. It follows that the behavior near the fixed points follows that of case (i) above. Instead at the non-classical fixed points we find $K = 4a^2/3 > 0$ and $A=0$ so case (ii) is relevant. The cycles look qualitatively like those shown in Fig.\ref{well}a. The phase shift and relative amplitude of state and control depend on the parameters.


\bibliography{../language,../Biology,../Ecology}

\end{document}